\documentclass[12pt]{article}

\usepackage{bbm}
\usepackage{amsmath}
\def\reptitle{Contractions of AdS brane algebra and  superGalileon Lagrangians }
\usepackage{amssymb} 

\catcode`@=11
\def\secteqno{\@addtoreset{equation}{section}%
\def\theequation{\thesection.\arabic{equation}}}
\catcode`@=12
\secteqno
\newcommand{\be}{\begin{equation}}
\newcommand{\ee}{\end{equation}}
\newcommand{\bea}{\begin{eqnarray}}
\newcommand{\eea}{\end{eqnarray}}
\newcommand{\bref}[1]{(\ref{#1})}
\newcommand{\nn}{\nonumber}
\newcommand{\A}{\alpha} \newcommand{\B}{\beta} \newcommand{\gam}{\gamma}
 \newcommand{\D}{\delta} 
\newcommand{\ep}{\epsilon} 
\newcommand{\T}{\theta} 
        \newcommand{\s}{\sigma}
          \newcommand{\w}{\omega}
          
\newcommand{\h}{\eta}           
           
\newcommand{\W}{\Omega}         
            \newcommand{\Sig}{\Sigma}

\newcommand{\ba}{\overline }
\def\6{\partial}
\def\7{\tilde}
\def\8{\hat}

\def\amu{a}\def\anu{b}\def\arho{c}\def\as{d}
\def\pa{\partial} 
 
\newcommand{\C}[1]{{\cal #1}}
\def\CL{{\cal L}}

\def\CP{{\cal P}}
\def\too{{\quad\to\quad}}
\def\sh{{\rm sh}}\def\ch{{\rm ch}}

\def\ebox#1#2{\vskip 2mm{\vbox{\hrule\hbox{\vrule\kern3pt\vbox{\kern3pt
         {\begin{eqnarray}#1\label{#2}\end{eqnarray}}
         \kern3pt}\kern3pt\vrule}\hrule}}\vskip 2mm}
\def\tbox{\vskip 2mm{\vbox{\hrule\hbox{\vrule\kern3pt\vbox{\kern3pt
         {{\hfill {\small ${}^{notebook\; kiyoshi}$} \\
         \large \bf ~~\reptitle}\\ } 
         \kern3pt}\kern3pt\vrule}\hrule}}\vskip 2mm}

\def\vs{\vskip 3mm}\def\={{\;=\;}}\def\+{{\;+\;}}
\begin{document}

\hfill {\today}

\begin{center} 

\vskip 15mm
{\Large \reptitle  }
\vskip 10mm 
{\large {Kiyoshi Kamimura} and {Seiji Onda}}\vskip 8mm
{\it 
  Department of Physics, Toho University Funabashi274-8510, Japan
 }
\vs
{\rm  kamimura@ph.sci.toho-u.ac.jp}
\date{\today}
\end{center}
\vskip 15mm

\begin{abstract}
We examine AdS Galileon Lagrangians using the method of non-linear realization. 
By contractions 1) flat curvature limit,  2) non-relativistic brane algebra limit
and 3) (1)+(2) limits we obtain DBI, Newton-Hoock and  Galilean Galileons respectively. We make clear how these Lagrangians appear  as invariant 4-forms and/or pseudo-invariant Wess-Zumino terms using Maurer-Cartan equations 
on the coset $G/SO(3,1)$. We show the equations of motion are written 
in terms of the MC forms only and explain why the inverse Higgs condition 
is obtained as the equation of motion for all cases.   

The supersymmetric extension is also examined using a supercoset  \\ $SU(2,2|1)/(SO(3,1)\times U(1))$ and  five WZ forms are constructed. They are reduced to the corresponding five Galileon WZ forms in the bosonic limit and are candidates for supersymmetric Galileon
action.   
\end{abstract}
\vs

{\it keywords: Galileon, non-linear realization, supersymmetry}  
\eject 

\section{Introduction}
Modifications of gravity using higher dimensional non-compact extra dimensions are 
important approaches to solve cosmological problems \cite{Randall:1999ee}.
The Galileons appear in such context \cite{Dvali:2000hr} and are interesting both in 
theoretical as well as phenomenological applications.  (See for example recent reviews \cite{Goon:2011xf}\cite{Curtright:2012gx}). 

It has been shown the Galileon actions are obtained based on the non-relativistic 3-brane 
algebra in 5 dimensions. It is the 5-dimensional Poincare algebra in which non-relativistic 
limit is taken in the transverse fifth direction.\footnote{The relativistic and non-relativistic 
brane actions was constructed using the non-linear realization of the brane algebras. 
See for example \cite{Gomis:2006xw}\cite{Gomis:2006wu}  and references therein. }
The Galileon appears as a Goldstone scalar field in the broken transverse direction of the 
3-brane and is satisfying second order equation of motion \cite{Goon:2012mu}. 
It has been clarified \cite{Goon:2012dy} that 5 possible forms of Galileon Lagrangians are 
the WZ Lagrangians constructed from closed and non-trivial 5-forms on the group manifold. 
It was also shown there is only one WZ Lagrangian for the DBI Galileon \cite{deRham:2010eu}\cite{Goon:2011qf} and conformal Galileon \cite{Nicolis:2008in}\cite{Goon:2012dy} theories. 

The Galileon Lagrangians are constructed \cite{Goon:2012dy} using the method of 
non-linear realization \cite{Coleman} for space-time symmetry algebras  \cite{Volkov}. 
The  Maurer-Cartan(MC) one forms on the coset $G/H$ are the building blocks of 
the Lagrangians. Here $G$ is the brane algebra and $H=SO(3,1)$ is the unbroken longitudinal 
Lorentz algebra of the brane. Galileons appear as the Goldstone mode with respect to the 
broken  transverse translation. The $G$-invariant Lagrangians are constructed from either $H$-invariant 
4 forms or pseudo-invariant   4-forms which are obtained from  closed and
 Chevalley-Eilenberg(CE)  non-trivial $H$-invariant 5 forms as the WZ Lagrangians 
\cite{Chevalley:1948zz,deAzcarraga:1998uy}. 

In this paper we restrict models of single Galileon and begin with the AdS algebra 
in 5 dimensions for AdS Galileon \cite{Burrage:2011bt}. This algebra allows three 
contractions giving four Galileon models;

$\star\;$  no-contraction gives AdS Galileon ,

$\star\;$ non-relativistic brane algebra limit gives Newton-Hoock(NH)  Galileons 

$\star\;$ flat curvature limit gives  Poincar\'e(DBI)  Galileons 

$\star\;$ non-relativistic and flat curvature  limits gives  Galilean Galileons. 
\begin{center}
\begin{tabular}{ccccc}
 &  &  & &  \\
 AdS & & ${\Longrightarrow}_{R\to\infty}$ & & Poincar\'e(DBI)   \\
 &  & & &  \\
 $\qquad \Downarrow_{\w\to\infty} $ & & & &
$\qquad\Downarrow_{\w\to\infty} $ \\
   &  & & &  \\
Newton-Hoock  &  & ${\Longrightarrow}_{R\to\infty}$  &  & Galilei  \\
\end{tabular}
\vs
Table 1:{~~~The contractions of AdS algebra}
\end{center}

They are examined in detail in   \cite{Goon:2012dy} and we extend the analysis to make 
clear systematically how five possible Lagrangians appear  either invariant 4 forms or
 pseudo-invariant Wess-Zumino terms depending on the contractions  
for all cases using Maurer-Cartan equations 
on the coset $G/SO(3,1)$. The equations of motion(EOM) are obtained by variations 
of the MC forms and we can express the EOM in terms of MC forms only, without 
using explicit parametrization of the coset. 

One purpose of this paper is to understand why the Galileons constructed in the
non-linear realization satisfy at most second order equations using Maurer-Cartan
equations for all four cases of Table 1. It comes from two facts; first is that the 
inverse Higgs condition, which eliminates the Goldstone boost vector variables 
in terms of the Galileon scalar,  is derived  as a EOM from the covariance.  
Second is the EOM for Galileon field is a (pullback of) sum of five invariant 4 forms 
and becomes at most second order differential equation for the Galileon scalar 
manifesting the Galileon property. 

Other is to apply the method of non-linear realization to  consider the supersymmetric 
extension of Galileon, which have been considered 
\cite{Khoury:2011da}\cite{Choudhury:2012yh}\cite{Koehn:2013hk} 
within superfield theory using, for 
example superfields. We start from a superalgebra $su(2,2|1)$ and taking Galilean limit 
we construct five closed  invariant 4-forms and five 5-forms which reduce to the ones 
of bosonic Galilean Galileon in the bosonic limit. In order to obtain these 
five candidates for the superGalileon WZ terms we need to enlarge the superalgebra 
with a fermionic and central extensions.
\vs

In section 2 we make a brief review of the NLR approach of the Galileons \cite{Goon:2012dy}
 clarifying how the WZ Lagrangians appear using the Maurer Cartan (MC) equation for every
 cases in Table 1. In section 3 we derive the equations of motion using with variation formula 
of the MC forms and derive the inverse Higgs condition for every cases.  The EOM for 
Galileons are sum of invariant 4-forms, which become second order for the Galileon scalar  
when the inverse Higgs condition is used.  In section 4 the conformal Galileon is discussed 
in the same context. In section 5 the supersymmetrization of the Galileon is considered. 
Summary and discussions are in the final section. There are three appendices for some 
useful formulas. Explicit forms of MC forms, for bosonic and super cases, 
are also presented by choosing coset parametrizations. 
  
 
\section{Relativistic and Non-relativistic Brane Algebras } 

In this section we give a reformulation of Galileon Lagrangians \cite{Goon:2012dy}
clarifying the "WZ" property in  using the MC equation.  We start with the AdS algebra in 
$d$ dimensions  and construct invariant $p$+1~-forms and closed and  invariant 
$p$+2~-forms to obtain candidates of  $p$-brane Lagrangians for Galileons.  

The AdS algebra in $d$ dimensions is $so(d$-1,2) and is written as
\bea
\left[P_{A},~P_{B}~\right]&=&-i~\frac{1}{R^2}\;M_{AB},\qquad 
\left[P_A,~M_{BC}\right]=-i\h_{A[B}~P_{C]},\nn\\
\left[M_{AB},~M_{CD}\right]&=&-i~\h_{B[C}~M_{AD]}+i~\h_{A[C}~M_{BD]},
\label{AdSD} 
\eea 
where $A,B=0,...,d-1$, $\h_{AB}=(-,+,...,+)$  and $R$ is the radius of AdS. 
In $R\to\infty$ limit it is contracted to the Poincare algebra.   In the presence of 
$p$-brane we split the space-time indices $A=0,...,d-1$ into the longitudinal directions 
${\amu}=0,1,...,p$ and transverse one $a'=p+1,...,d-1$. The Newton Hoock(NH)  algebra 
\cite{Bacry:1968zf} for non-relativistic brane \cite{Brugues:2006yd} is that in which 
the light velocity goes to infinity in the transverse directions.  We get it by a contraction 
of the AdS algebra \bref{AdSD} using a rescaling  
\be
P_{a'}\to {\w} P_{a'}, \qquad M_{{\amu} {a'}}\to \w M_{{\amu} a'} 
\label{Galrescale}\ee 
and $\w\to\infty$.  
In this paper we apply it to single Galileon models and 
we restrict one transverse direction, i.e.  
$p$-brane in $d=p+2$ dimensions. Writing the single transverse index as $a'=\pi$
and the boost generators in the transverse direction as $M_{a\pi}=B_a$ the algebra becomes  
\bea
\left[P_{{\amu}},~P_{{\anu}}~\right]&=&-i~ \frac{1}{R^2}\;M_{{\amu}{\anu}},\quad
\left[P_{{\amu}},~P_{\pi}~\right]=-\frac{i}{R^2}\;B_{{\amu}}, \quad 
\left[P_{\amu},~M_{{\arho}{\as}}\right]=-i~\h_{{\amu}[{\arho}}~P_{{\as}]},
\nn\\
\left[B_{{\amu}},~P_{{\anu}}~\right]&=&i~ \h_{{\amu}{\anu}}~P_{\pi},\qquad
\left[B_{{\amu}},~P_{\pi}~\right]=-\frac{i}{\w^{2}}\,P_{{\amu}},\quad
\left[B_{{\amu}},~B_{{\anu}}\right]=\frac{i}{\w^2}~M_{{\amu}{\anu}},
\nn\\
\left[B_{{\amu}},~M_{{\arho}{\as}}~\right]&=&-i~\h_{{\amu}[{\arho}}~B_{{\as}]},\quad
\left[M_{{\amu}{\anu}},~M_{{\arho}{\as}}\right]=-i~\h_{{\anu}[{\arho}}~M_{{\amu}{\as}]}+
i~\h_{{\amu}[{\arho}}~M_{{\anu}{\as}]}.
\label{AdS23}
\eea
In the $R\to\infty$  limit it becomes the Poincar\'e (DBI) brane algebra and  
in the $\w\to\infty$ it goes to  NH brane algebra. 
Taking both   $\w\to\infty$ and  $R\to\infty$  limits it becomes Galilean
brane algebra (Galileon algebra). We consider these four cases by comparison.  
(Table 1)
 
\vs
Taking the AdS algebra  $G$ in  \bref{AdS23} and the
stability group $H$ as the longitudinal Lorentz algebra $so(p,1)$  we describe the system using a coset 
$G/H$.   The Maurer-Cartan(MC) form $\W$ is introduced by
\be
\W=-ig^{-1}dg=G_AL^A=P_{\amu} L_P^{\amu}+\frac12\,M_{{\amu}{\anu}}\,L^{{\amu}{\anu}}+P_\pi L^\pi +\,B_{{\amu}}\,L_B^{{\amu}} ,\qquad g\in G/H.
\label{MC0}\ee 
Using the first expression of $\W=-ig^{-1}dg$ it holds identically the MC equation 
\be d\W+i\W\wedge\W=0. 
\label{MCW}\ee
Using the second expression of \bref{MC0}, $\W=G_AL^A$, and for algebra $[G_A,G_B]=if^C_{AB}G_C$ \bref{MCW} gives MC equation for the component one forms 
$L^A$'s as 
\be
dL^A+\frac12 f^A_{BC}L^C\wedge L^B=0.
\ee
For the AdS algebra \bref{AdS23} the  MC equation becomes\footnote{
We often abbreviate $"\wedge"$ symbol for wedge products.}
\bea
dL_P^{\amu}&+&L_P^{\arho}{L_{\arho}}^{\amu}+\frac{1}{\w^2}{L_B^{{\amu}}}L^\pi=0,
\nn\\
dL^\pi&+&L_P^{\arho}{L_{B\arho}}=0,
\nn\\
dL^{{\amu} {\anu}}&+&L^{{\amu} {\arho}}{L_{\arho}}^{\anu}-
\frac{1}{\w^2}L_B^{{\amu}}{L_B^{{\anu}}}+
\frac{1}{R^2}L_P^{\amu} L_P^{\anu}=0,
\nn\\
dL_B^{{\amu}}&+&L^{{\amu} {\arho}}{L_{B\arho}}+
\frac{1}{R^2}L_P^{\amu} L^\pi=0.\label{AdSw}
\eea
The consistency(integrability) of the set of MC equations \bref{AdSw} is equivalent to 
holds the Jacobi identities of the algebra  \bref{AdS23}. (See Appendix A where some useful formulas are summarized.)

In the non-linear realization on the coset $G/H$ the coset elements $g$'s are parametrized by coset coordinates  $Z^M$. Under infinitesimal global $G$ transformations the coset element $g$ transforms as 
\be
g(Z)\too g'=g_\ep\,g(Z)\,h^{-1}(\ep,Z)=g(Z'),\qquad  g_\ep\in G,\qquad h\in H,
\label{inftransg}\ee
where $h(\ep,Z)$ is the compensating local $H$ transformation so that $g'$ becomes a coset element. 
The MC form $\W$ transforms under $G$ as 
\be
\W\too \W'=h\,\W\,h^{-1}-ih\,d\,h^{-1}.
\ee
Since the last term belongs to the subalgebra $H$  one forms $L$'s associated with $G/H$ transform as $H$ 
covariants and  $L$'s associated with $H$ transform as $H$ gauge connections. 
Now $L^\pi$ is a scalar, $L_P^{\amu}$ and $L_B^{{\amu}}$ are  $so(p,1)$ vectors and $L^{{\amu}{\anu}}$ is a gauge connection of the $so(p,1)$ under $G$ transformations.  
The $G$-invariant $p$-brane action can be constructed from  local $H$ (thus $so(p,1)$ Lorentz) invariant $p$+1 forms.
In addition when closed and  local $H$-invariant $p$+2 forms exist and 
are CE non-trivial, that is they are not written as "$d$" of some $p$+1 forms  from $L$'s,     
we can construct WZ Lagrangians that is pseudo invariant under $G$ transformations 
from them \cite{Chevalley:1948zz}\cite{deAzcarraga:1998uy}. 

\vs
We can construct $H$-invariant $p$+1  forms $K_q, (q=0,1,...,p+1) $ from wedge products 
of $q$  vectors $L_P^{\amu}$ and  $(p+1-q)$  vectors $L_B^{{\amu}}$,
\bea
K_{q}&\equiv&\ep_{{\amu}_0...{\amu}_{q-1}{\amu}_{q}...{\amu}_p }\,L_P^{{\amu}_0}...
L_P^{{\amu}_{q-1}}\,L_B^{{\amu}_{q}}...L_B^{{\amu}_{p}},\qquad (q=0,...,p+1),
\label{defKq}\eea 
where $\ep_{{\amu}_0...{\amu}_p}$ is the Levi-Civita tensor with $\ep_{0...p}=+1$. 
 $K_q$'s are only possible non-trivial invariant $p$+1  forms constructed from wedge 
product of the MC one forms.\footnote{ 
There is a $H$-invariant 2-form $(\h_{ab}L_P^{\amu} \wedge  L_B^b=-dL^\pi)$ contracted using $\h_{ab}$. However a possible 4-form  is exact 
$(\h_{ab}L_P^{\amu}\wedge  L_B^b)\wedge(\h_{cd}L_P^{c}\wedge  L_B^d)=dL^\pi\wedge dL^\pi$ and is not used for three brane Lagrangian in 4 dimensions.}  
Taking wedge product with $H$-scalar $L^\pi$ we define  $H$-invariant $p$+2   forms 
$\W_q$ as  
\bea
 \W_q=K_{q}\,L^\pi,\qquad (q=0,...,p+1).
\label{defWq}
\eea
They are only possible $H$-invariant $p+2$ forms using wedge products of $L$'s. 

Using the MC equations \bref{AdSw}  they are related by 
\bea
dK_q&=&
-\frac{(-)^{p}}{\w^2}\,q\, \W_{q-1}-
\frac{(-)^{p}}{R^2}(p+1-q)\,\W_{q+1}, 
\label{dKeqW}\eea
and $\W_q$'s are closed,
\bea
d  \W_q &=&dK_{q}\,L^\pi+(-)^{p+1}\,K_q\,(-L_{Pa}{L_B^{a}})=0. 
\label{dWq} \eea
The first term of \bref{dWq} vanishes since $dK_{q}$ includes $L^\pi$ as in \bref{dKeqW}. In the second term
either $L_P^{a}$ or $L_B^{{a}}$ exists in $K_q$ for every ${a}$ then it vanishes. 
Thus $\W_q$'s are $H$-invariant and closed $p$+2   forms.

These $K_q$ and $\W_q$ are used for constructing the Galileon Lagrangians
by taking pull back to the $p$-brane world volume \cite{Goon:2012dy}.   
$K_q$ is a possible candidate of the $G$-invariant Lagrangian when it is not closed.
 $\W_q$ is used to construct $G$-pseudo-invariant WZ Lagrangian 
 as $\W_q=d\CL_q^{WZ}$ 
when $\W_q$ is  CE non-trivial, i.e. it is not expressed in a form $d \Lambda^{p+1}$ with
some $H$-invariant $p+1$ form  $\Lambda^{p+1}$ constructed from $L$'s. Since  only possible non-trivial invariant $p$+1  forms are $K_q$'s~  $\Lambda^{p+1}$ is a linear combination of  $K_q$'s for CE-trivial $\W_q$. 
Eq.\bref{dKeqW} tells the CE 
cohomological structure 
and numbers of non-closed $K_q$'s 
and non-trivial $\W_q$'s, i.e. numbers of  invariant and WZ Lagrangians.  
\bref{dKeqW} is written in a matrix form as
\bea
\begin{pmatrix}dK_0 \cr dK_1 \cr \vdots \cr dK_{p}\cr dK_{p+1}
\end {pmatrix}=-(-)^p
\begin{pmatrix}0 &\frac{p+1}{R^2}&   \cr
\frac{1}{\w^2}&0&\frac{p}{R^2}&& \cr
&\frac{2}{\w^2}&0&\frac{p-1}{R^2}& \cr
 &&\ddots&\ddots&\ddots\cr
&&&\frac{p}{\w^2}&0&\frac{1}{R^2}& \cr
&&&&\frac{p+1}{\w^2}&0\cr
\end {pmatrix}
\begin{pmatrix}\W_0\cr\W_1\cr\vdots\cr \W_{p}\cr \W_{p+1}
\end {pmatrix}
\label{dKeqWM}\eea
Let $r$ is the {\it rank} of the $(p+2)\times(p+2)$ matrix $M$ appearing in \bref{dKeqWM}, $r$ of $\W_q$ 
are expressed in terms of $dK$'s then are CE trivial.  The remaining $(p+2-r)$ of $\W_q$ 
are CE non-trivial and the number of WZ Lagrangians is  $(p+2-r)$. 
It also tells there are  $(p+2-r)$ independent linear combinations of $dK$'s that vanish
identically. There remain $r$ linear combinations of $K$'s that are not closed which
gives $r$ $G$-invariant Lagrangians.    
In total there are $p+2$ possible Lagrangians,   $r$  $G$-invariant Lagrangians
and $(p+2-r)$ WZ Lagrangians. 

The number of  WZ Lagrangians $(p+2-r)$ is determined by the rank $r$ of the
matrix  $M$  in \bref{dKeqWM} and it depends on how the algebra is contracted.
For the four contractions  in the Table 1 ranks $r$ of the matrix $M$ are 
\bea
{\rm  Galilean~Galileon~~} (\w\to\infty, R\to\infty) &:&r=0,
\nn\\ 
{\rm  DBI ~Galileon~~} (\w=1, R\to\infty) &:&r=p+1,
\nn\\ 
{\rm  NH ~ Galileon~ ~~} (\w\to\infty, R=1) &:&r=p+1,
\nn\\ 
{\rm  AdS ~ Galileon~ (for~odd~}p)~~ (\w=1) &:&r=p+1,
\nn\\ 
{\rm ~ (for~even~}p)~~ (\w=1) &:&r=p+2.
\nn\eea
We will examine the possible Lagrangians for these cases in some detail.  
\vs

{\underline{ 1) Galilean brane (Galileon)}}  [$\w\to\infty, R\to\infty$] 
\vs
 \bref{dKeqWM} tells $M=0$ and $r=0$, all of the invariant $p$+1  forms 
$K_{q}$'s are closed in this limit $\w\to\infty, R\to\infty$,
\be
dK_q=0,\qquad  (q=0,1,...,p+1).
\label{GalileondK}\ee
If $K_{q} $'s  are used in the Lagrangian they are surface terms. 
On the other hand closed $p$+2   forms $\W_{q}, (q=0,1,...,p+1)$ are CE non-trivial
and are used as WZ $p$+2   forms. 
There are $p$+2   WZ Lagrangians  $\CL^{WZ}_q,(q=0,1,...,p+1)$ satisfying $\W_q=d\CL^{WZ}_q$  and  $\CL^{WZ}_q$'s are pseudo-invariant $p$+2   Galileon Lagrangians \cite{Goon:2012dy}. Due to  CE non-triviality of $\W_q$ the WZ Lagrangians $\CL^{WZ}_q$ are not expressed using MC forms. It requires 
 coset coordinates to write down the   WZ Lagrangians $\CL^{WZ}_q$ explicitly.
  
\vs

{\underline{ 2) Poincar\'e brane (DBI Galileon)}}   [$\w=1, R\to\infty$] , 
\vs
In this case  \bref{dKeqWM} shows
\be
dK_{0}=0 ,\qquad \W_{q}=-\frac{(-)^p}{(q+1)}\,dK_{q+1}, \qquad (q=0,1,...,p). 
\label{eq2.11}\ee
$K_0$ is closed and is a surface term. 
$K_q, (q=1,...,p+1) $'s are $H$-invariant $p$+1  forms and are used as the Lagrangians 
(Lovelock invariants) \cite{Lovelock:1971yv}.  
$\W_{q}, (q=0,1,...,p) $'s are  CE trivial 
since they are given as $"d"$ of  MC forms $K_{q+1}$ as in \bref{eq2.11}. 
The closed $p$+2   form $\W_{p+1}=K_{p+1}L^\pi$ is however non-trivial and is used
as the WZ Lagrangian $\CL^{WZ}_{p+1}$ satisfying 
$\W_{p+1}=K_{p+1}L^\pi =d\CL^{WZ}_{p+1}$.
The only one WZ Lagrangian  $\CL^{WZ}_{p+1}$  is the tadpole term \cite{Goon:2012dy} . 

\vs

{\underline{ 3)  NH  Galileon} }  [$\w\to\infty, R$ finite].

\vs
In this case \bref{dKeqWM} gives 
\be 
\W_{q+1}=-\frac{{R^2}(-)^{p}}{(p-q+1)}\,dK_q , 
\qquad  (q=0,...,p), \qquad    dK_{p+1}=0. 
\label{dKeqW4}
\ee
$K_{p+1}$ is closed and gives a surface term (cosmological constant).   
$\W_q, (q=1,...,p+1)$ are written in terms of $dK_{q-1}$ and are 
CE trivial as in \bref{dKeqW4}.  There is only one CE non-trivial closed $p$+2 form $\W_0$ that is used 
as the WZ $p$+2   form to construct WZ Lagrangian $\W_0=d\CL^{WZ}_0$. 
\vs
{\underline{4) AdS   Galileon}   [$\w=1, R$ finite],  
\vs
In this case the rank of the matrix $M$ in \bref{dKeqWM} is $p+1$ for odd $p$ and 
$p+2$ for even $p$.  
Eq.\bref{dKeqWM}  is more explicitly,
\bea
dK_0&=&-(-)^p\left(\qquad \qquad  \frac{p+1}{R^2}\,\W_1\right), \qquad 
dK_1=-(-)^p\left(\,\W_0+ \frac{p}{R^2}\,\W_2\right), 
\nn\\ 
dK_2&=&-(-)^p\left(\,2\W_1 +\frac{p-1}{R^2}\,\W_3\right), \qquad 
\hskip 15mm..... \nn\\ && .....\hskip 55mm
dK_{p-1}=-(-)^p\left(\,(p-1)\W_{p-2}+ \frac{2}{R^2}\,\W_{p}\right), \nn\\ 
dK_p&=&-(-)^p\left(\,p\W_{p-1}+ \frac{1}{R^2}\,\W_{p+1}\right), \qquad 
dK_{p+1}=-(-)^p\quad (p+1)\W_p.
\eea
The first equation means $\W_1$ is proportional to $dK_0$ and is 
CE trivial, the third one tells
$\W_3$ is also 
CE trivial. Then all $\W_{odd}$ are  CE trivial. Similarly  
starting  from the last equation $\W_p, \W_{p-2},...$ are CE trivial. 
When $p$ is even all $\W_q, (q=0,...,p+1)$ can be expressed as a linear combination of
$dK_j$ then are CE trivial. 
On the other hand for odd $p$ only $\W_q, (q=1,3,...,p)$ can be expressed as linear 
combinations of $dK_{even}$ with a closure relation 
\be
d\left[\sum_{i=0}^{\frac{p+1}2}\frac{(-)^i}{R^{2i}}\frac{(p+1)!!}{(2i)!!(p+1-2i)!!}\,K_{2i}\right]=0.
\label{closedi}\ee 
$\W_{q}, (q=0,2,...,p+1)$ are expressed using $dK_{odd}$'s and one of $\W_{even}$. For example
$K_q, (q=0,...,p)$ can be taken as independent non-trivial invariant $p$+1  forms 
and $\W_{p+1}$ is used to construct the WZ Lagrangian   $\CL^{WZ}_{p+1}$. 
$K_{p+1}$ is a linear combination of other $K_{even}$ up to closed form due to \bref{closedi} 
and  $\W_{q},(q=0,...,p)$ are expressed in terms of $dK_q$'s and $\W_{p+1}$. 

\vs
In summary Galileon Lagrangians are given by taking pullback of these forms,\footnote{
Lagrangians are pullbacks of forms to the world-volume. The pullback notation, $L^{a}\rightarrow L^{a*}$, etc., is omitted for simplicity.}  \\
1) Galilean brane,  $[\w\to\infty,\; R\to\infty],$ 
\be
\CL^{Gal}=\sum_{q=0}^{p+1}\,b^q\,\CL^{WZ}_{q}.
\label{LagGal}\ee 
2)  {Poincare} brane (DBI),  $[\w=1,\; R\to\infty],$ 
\be
\CL^{Poincare}=\sum_{q=1}^{p+1}\,c^q\,K_q\,+b^{p+1}\,\CL^{WZ}_{p+1}.
\label{LagPoin}\ee
3)  NH Galileon ,  $[\w\to\infty, \; R$ finite],
\be
\CL^{NH}=\sum_{q=0}^{p}\,c^q\,K_q\,+b^{0}\,\CL^{WZ}_{0}.
\label{LagNH}\ee
4) AdS  Galileon ,   $[\w=1,\; R$ finite],
\bea
\CL^{AdS}&=&\sum_{q=0}^{p+1}\,c^q\,K_q,\hskip 30mm ({\rm for\; even\;} p), 
\\
\CL^{AdS}&=&\sum_{q=0}^{p}\,c^q\,K_q\,+b^{p+1}\,\CL^{WZ}_{p+1},\qquad ({\rm for\; odd\;} p).
\label{LagAdS}\eea
Each Lagrangian has $p$+2 independent terms, apart from surface terms.
They are formally written as
\be
\CL^{tot}=\sum_{q=0}^{p+1}\left[c^qK_q+b^q\CL^{WZ}_{q}\right],
\label{Lagtot}\ee
where only $p$+2   of coefficients $c^q$ and $b^q$ are non-vanishing depending on the cases 
as above. 
\begin{center}
\begin{tabular}{|c c c||c c||cc||cc||cc|}
  \hline
&  Gal &  & DBI  & & NH   &  & AdS  &even $p$ &AdS  & odd $p$  \\
  \hline  
&&   $\CL^{WZ}_{0}$  &  & & $K_{0}$&$\CL^{WZ}_{0}$ & $K_{0}$& & $K_{0}$& $\circ$ \\
&&  $\CL^{WZ}_{1}$  & $K_1$&  & $K_1$& & $K_1$& & $K_1$&  \\
&  & $\CL^{WZ}_{2}$  & $K_2$& & $K_2$&& $K_2$& & $K_2$& $\circ$  \\
&  & $\vdots $  &  $\vdots $& &  $\vdots $&  &  $\vdots $& & $\vdots $ & \\
&  & $\CL^{WZ}_{p-1}$  & $K_{p-1}$ & & $K_{p-1}$ & & $K_{p-1}$ & &$K_{p-1}$ & $\circ$ \\
&  & $\CL^{WZ}_{p}$  & $K_p $ & & $K_p $ & & $K_p $& & $K_p $&  \\
&  & $\CL^{WZ}_{p+1}$  & $K_{p+1}$ &$\CL^{WZ}_{p+1}$&  & &$K_{p+1}$ & & $\circ$ &$\CL^{WZ}_{p+1}$  \\
  \hline
&0&p+2&p+1&1&p+1&1&p+2&0&p+1&1\\
\hline
\end{tabular}
\vs
Table 2:{~~Possible invariant and WZ Lagrangians. 
In the last line numbers of invariant Lagrangian (left) and number of WZ Lagrangian (right) are tabulated. There are ambiguity in AdS(odd $p$) case. Terms indicated by $\circ$ could appear but are dependent.}
\end{center}

\section{Equations of motion} 

In this section we derive the equations of motion (EOM) by taking the variation of 
the Lagrangians with respect to the coordinates of the coset.   
Since the Lagrangians are constructed from MC forms not only variations of the invariant Lagrangians but also those of the  WZ Lagrangians are expressed
in terms of MC forms. Especially we get the inverse Higgs condition $L^\pi=0$ 
as a result of  the EOM for all cases. 

Variations of MC forms, under any  variation of the coset coordinates 
$Z^M$, is given by \bref{deltaLIF},
\be
\D L^A=d[\D Z]^A +f^A_{BC}\;{L}^C{[\D Z}]^B
\label{VarL}\ee
where  $f^A_{BC}$ is the structure constants of the algebra and
$[\D Z]^A $ is defined by replacing $dZ^M$ with $\D Z^M$ in the MC form $L^A$,
\be
[\D Z]^A\equiv \D Z^M{L_M}^A\quad {\rm for}\quad 
L^A =d Z^M{L_M}^A.
\ee
An advantage of using \bref{VarL} is that it does not depend on how the coset is parametrized.
In the present case 
the coset coordinates $Z^M$ are $x^{\amu}, \pi, v^{\amu}$ associating to the 
$G/H$ generators, $P_{\amu}, P_\pi, B_{{\amu}}$ respectively. 
$[\D Z]_P^{\amu}, [\D Z]^\pi,
 [\D Z]_B^{{\amu}} $ are $L_P^{\amu}, L^\pi,L_B^{{\amu}}$ in which $dx^{\amu}, d\pi, dv^{\amu}$ are replaced by $\D x^{\amu}, \D\pi, \D v^{\amu}$ respectively. 
\bref{VarL} becomes 
\bea
\D L_P^{\amu}&=& d[\D Z]_P^{\amu} +L_P^{\arho}{[\D Z]_{\arho}}^{\amu}- 
[\D Z]_P^{\arho}{L_{\arho}}^{\amu}+\frac{1}{\w^2}{L_B}^{{\amu}}[\D Z]^\pi
-\frac{1}{\w^2}{[\D Z]}_B^{{\amu}}L^\pi,
\nn\\
\D L^\pi&=&d[\D Z]^\pi+L_P^{\arho}{[\D Z]_{B\arho}}-[\D Z]_P^{\arho}{L_{B\arho}},
\nn\\
\D L^{{\amu} {\anu}}&=& d[\D Z]^{{\amu} {\anu}}+L^{[{\amu} {\arho}}{[\D Z]_{\arho}}^{{\anu}]}
-\frac{1}{\w^2}L_B^{[{\amu}}{[\D Z]}_B^{{\anu}]}+
\frac{1}{R^2}L_P^{[{\amu}}[\D Z]_P^{{\anu}]},
\nn\\
\D L_B^{{\amu}}&=& d[\D Z]_B^{{\amu} }+L^{{\amu} {\arho}}{[\D Z]_{B\arho}}-
[\D Z]^{{\amu} {\arho}}{L_{B\arho}}+\frac{1}{R^2}L_P^{\amu} [\D Z]^\pi-\frac{1}{R^2}[\D Z]_P^{\amu} L^\pi.
\label{AdSw12v}
\eea
 Using it we compute variations of $K_q$ and $\CL^{WZ}_q$ under general variations
\bref{AdSw12v}.
For $K_q$, apart from exact forms,
\bea
\D K_{q}&=&
-\left\{
\frac{1}{\w^2}
q(p-q+2)\, \ep_{{\amu}_0...{\amu}_{q-2}{\amu}_{q-1}...{\amu}_p }\,
L_P^{{\amu}_0}...L_P^{{\amu}_{q-2}}\, L_B^{{\amu}_{q-1}}...L_B^{{\amu}_{p-1}}\,
\right.\nn\\&&\left.+\frac{1}{R^2}
(p-q)(p-q+1)\, \ep_{{\amu}_0...{\amu}_{q}{\amu}_{q+1}...{\amu}_p }\,L_P^{{\amu}_0}...
L_P^{{\amu}_{q}} \, L_B^{{\amu}_{q+1}}...L_B^{{\amu}_{p-1}}\right\}L^\pi [\D Z]_B^{{\amu}_p}
\nn\\ &-&
\left\{ \frac{1}{\w^2}q(q-1)\, \ep_{{\amu}_0...{\amu}_{q-3}{\amu}_{q-2}...{\amu}_p }\,L_P^{{\amu}_0}...
L_P^{{\amu}_{q-3}}\,L_B^{{\amu}_{q-2}}\,...L_B^{{\amu}_{p-1}} 
\right.\nn\\&&\left.+\frac{1}{R^2}(q+1)(p-q+1)\, \ep_{{\amu}_0...{\amu}_{q-1}{\amu}_{q}...{\amu}_p }\,L_P^{{\amu}_0}..
.\,L_P^{{\amu}_{q-1}}L_B^{{\amu}_{q}}...L_B^{{\amu}_{p-1}}\right\}L^\pi\,[\D Z]_P^{{\amu}_{p}},
\nn\\ 
&+&\left\{\frac{q}{\w^2}\,K_{q-1} \,
+\frac{(p-q+1)}{R^2}\, K_{q+1}\right\} [\D Z]^{\pi}. 
\label{delKq}\eea
Since the WZ Lagrangian $\CL^{WZ}_q$ is defined from the closed form as 
$\W_q=d\CL^{WZ}_q$ the variation of  $\CL^{WZ}_q$ is determined 
from that of  $\W_q$.  
Actually $\D\W_q$ is written in an exact form  
and  $\D\CL^{WZ}_q$ is read from $\D\W_q=d[\D \CL^{WZ}_q]$,  up to closed form, as
\bea
\D\CL^{WZ}_q&=&(-)^p
(p-q+1)\, \ep_{{\amu}_0...{\amu}_{q-1}{\amu}_{q}...{\amu}_p }\,L_P^{{\amu}_0}...L_P^{{\amu}_{q-1}}\,
 L_B^{{\amu}_{q}}...L_B^{{\amu}_{p-1}}\,L^\pi\,[\D Z]_B^{{\amu}_p}
 \nn\\&+&
(-)^{p}q\, \ep_{{\amu}_0...{\amu}_{q-2}{\amu}_{q-1}...{\amu}_p }\,L_P^{{\amu}_0}...L_P^{{\amu}_{q-2}}\,
L_B^{{\amu}_{q-1}}...L_B^{{\amu}_{p-1}}\,L^\pi \,[\D Z]_P^{{\amu}_{p}} 
\nn\\  &-&(-)^{p}K_{q}\,[\D Z]^{\pi}.
\label{delWZL}\eea
Note there appears no $[\D Z]^{{\amu}{\anu}}$ term in $\D K_{q}$ and $\D\CL^{WZ}_q$
due to local $SO(p,1)$ invariance. 

In deriving the Euler-Lagrange(EL) equations we take variation of the Lagrangians 
with respect to the coset coordinates $\{x^{\amu}, \pi, v^{\amu}\}$ associating to the 
$G/H$ generators, $\{P_{\amu}, P_\pi, B_{{\amu}}\}$. In above $\{[\D Z]_P^{\amu}, [\D Z]^\pi,
 [\D Z]_B^{{\amu}}\} $ are $\{L^{\amu}, L^\pi,L_B^{{\amu}}\}$ in which $\{dx^{\amu}, d\pi, dv^{\amu}\}$ are
replaced by $\{\D x^{\amu}, \D\pi, \D v^{\amu}\}$ respectively. Since  $\{x^{\amu}, \pi, v^{\amu}\}$ 
are a parametrization of the coset  $\{[\D Z]_P^{\amu}, [\D Z]^\pi, [\D Z]_B^{{\amu}}\} $ and  
$\{\D x^{\amu}, \D\pi, \D v^{\amu}\}$ are linearly related by a non-singular matrix. 
The Euler-Lagrange equations are thus coefficients of   $\{[\D Z]_P^{\amu}, [\D Z]^\pi,
 [\D Z]_B^{{\amu}}\} $ in the variations of the Lagrangians \bref{delKq} and \bref{delWZL}.
 It is important to notice the EL equations are written in terms of (pullback of)
exterior products of $p$+1 MC 1-forms.  It is contrasted with the fact that 
 explicit forms of the WZ Lagrangians can not be written in terms of MC forms but require  coset coordinates 
 due to CE non-triviality of WZ $p$+2 -forms $\W_q$.  
 
 Now the Lagrangians \bref{LagPoin}-\bref{LagAdS} are linear combinations of $K_q$ and 
$\CL^{WZ}_q$ as \bref{Lagtot} coefficients of  $[\D Z]_B^{{\amu}}$ in $\D K_q$ and $\D\CL^{WZ}_q$ have common factor $L^\pi$ and
 \be
\D\CL=(p{\rm -form})_{{\amu}}\wedge L^\pi\,[\D Z]_B^{{\amu}}=0.
\label{ELvmu}\ee
It gives $p$+1  $({\amu}=0,...,p)$ components independent EL equations and is solved by 
\be
L^\pi=0, 
\label{IHcond}\ee
which is a one-form equation in $p$+1  dimensions thus including $p$+1  independent 
components.
The equation \bref{IHcond} is known as the inverse Higgs condition\cite{Ivanov:1975zq}
 either obtained as EOM or imposed in models using non-linear realization. 
In the present case it is derived  as the EL equations 
from the variation of the boost coordinates $v^{\amu}$.  
We can understand the reason why  $L^\pi$  appears in the variation of $p$+1  forms 
$K_q$ in  \bref{delKq}  and $\CL^{WZ}_q$ in \bref{delWZL} as the common factor. Each coefficients of $[\D Z]_B^{{\amu}}$ 
in $\D K_q$ and $\D\CL^{WZ}_q$ 
is $p$+1  form and $H$-vector with the index ${\amu}$. To construct such term  
from the MC forms, $L^{\amu},L^{{\amu}\pi}$ and $L^\pi$, it is required to use one scalar
one form  $L^\pi$ in addition to $p$ of $L_P^{\amu},L_B^{{\amu}}$ from the $H$-covariance.   

The same argument is applied in the variation of $x^{\amu}$, 
the coefficients of $[\D Z]_P^{{\amu}} $ in the variations \bref{delKq} and \bref{delWZL} have the common factor $L^\pi$.
Then it does not give independent EOM. 
The fact that they are dependent is a reflection of the diffeomorphism invariance which is
manifestly assured in the differential form description. We could take the 
coset parameters $x^{\amu}$ associating to $P_{\amu}$ non dynamically. 

The  $[\D Z]^{\pi} $ term in the variation of the total Lagrangian $\CL^{tot}$
in \bref{Lagtot} gives
the EL equation %
\bea
\sum_{q=0}^{p+1}\left[ \,c^q\,
\left(\frac{q}{\w^2}\,K_{q-1} +\frac{(p-q+1)}{R^2}\, K_{q+1}\right)\,+\,b^q\,
 (-)^{p}K_{q}  \right] \,=0,
\label{piEOM}\eea 
where $p$+2   of coefficients $c^q$ and $b^q$ vanish identically depending on the cases 
in \bref{LagPoin}-\bref{LagAdS}. It is noted that the EOM 
\bref{piEOM} is a linear combination of all $K_q,(q=0,...,p+1)$ for every cases. 
 For example for DBI case ($R\to\infty$) \bref{piEOM} is 
 \bea
\,c^1\,K_0+2\,c^2\,K_1+...+(p+1)\,c^{p+1}\,K_p\,+\,b^{p+1}\,
 (-)^{p}K_{q+1} \,=0.
\label{piEOMDBI}\eea 

The inverse Higgs condition \bref{IHcond}, by taking pullback to the $p$-brane 
world volume, is a set of  algebraic equations determining the
coset coordinates $v^{\amu}$ associating to $B_{{\amu}}$ 
in terms of $\pi$, the
coset coordinate associating to $P_{\pi}$,  and its first order derivatives, 
(see Appendix B for explicit forms), 
\be
L^\pi=0, \quad\to\quad v^{\amu}=v^{\amu}(\pi,\pa\pi).
\label{IHcondsol}\ee
Using it in \bref{piEOM} it becomes EOM for $\pi$. 
Since $L$'s includes $v^{\amu}$ at most first order derivatives,  
the EOM  \bref{piEOM}, after eliminating  $v^{\amu}$ in terms of $\pi$  
is at most second order differential equation of $\pi$. 
Then $\pi$ is the Galileon field verifying
second order EOM. 
Note we have not used any particular choice of the coset parametrization and above
results are general ones. 
 
One could use  \bref{IHcondsol} in the Lagrangian 
$\CL^{tot}(x^{\amu},v^{\amu},\pi)$ in \bref{Lagtot} to define effective Lagrangian 
$\CL^{\pi}(x^{\amu},\pi)$. 
Although it depends on the second order derivatives of $\pi$ the 
EL equation for $\pi$ is not higher order but remains to be  
second order one. It is equivalent to 
one obtained from  \bref{piEOM} using  \bref{IHcondsol}
since $v^{\amu}$ is solved 
algebraically as in \bref{IHcondsol}.     

In Appendix B we solve the IH equation \bref{IHcondsol} in a parametrization of the coset
 $G/H$ and find expressions of MC forms in each cases. 

\section{Conformal Galileon}

In above we have considered  AdS and its contracted Galileons. 
Here we consider briefly the conformal Galileons \cite{Nicolis:2008in}
 in a similar manner as the previous sections.  
The conformal algebra in 4-dimensions is so(4,2) 
and is 
\bea
\left[P_{{\amu}},~K_{\anu}\right]&=&+2i~M_{{\amu}{\anu}}-2i~\h_{{\amu}{\anu}}\,D,
\quad \left[P_{{\amu}},~D~\right]=-i~P_{{\amu}}, 
\quad \left[K_{{\amu}},~D~\right]=i~K_{{\amu}},
\nn\\
 \left[M_{{\amu}{\anu}},~D~\right]&=&0, \qquad 
\left[M_{{\amu}{\anu}},~M_{cd}\right]=-i~\h_{{\anu}[c}~M_{{\amu}d]}+
i~\h_{{\amu}[\rho}~M_{{\anu}\s]},
\nn\\
\left[P_{\amu},~M_{cd}\right]&=&-i~\h_{{\amu}[c}~P_{d]}, \qquad 
\left[K_{\amu},~M_{cd}\right]=-i~\h_{{\amu}[c}~K_{d]}. \qquad 
\label{conf4}
\eea
The MC form is
\be \W=
P_{{\amu}}L_P^{{\amu}}+K_{\amu}\,L_K^{{\amu}}+D\,L_D+\frac12M_{{\amu}{\anu}}L^{{\amu}{\anu}},
\ee
and the MC equation is
\bea
dL^{{\amu} {\anu}}+L^{{\amu} c}{L_c}^{\anu}-2\,L_P^{[{\amu}}{L_K}^{{\anu}]}&=&0,\qquad 
dL_D+2L_P^{c}{L_{Kc}}=0.
\nn\\
dL_P^{{\amu}}+L^{{\amu} c}L_{{Pc}}+L_P^{{\amu}}{L_D}&=&0,\qquad 
dL_K^{{\amu}}+L^{{\amu} c}L_{{Kc}}-L_K^{{\amu}}{L_D}=0.
\eea
\vs
When we consider a coset $G/H=$SO(4,2)/SO(3,1)=Conf/Lorentz the 3-brane actions
are SO(3,1) invariant 4-forms or WZ action obtained from SO(3,1) invariant and closed 5-forms.
$H$-invariant 4-forms constructed from the MC one forms are
\bea
K_0&=&\ep_{{\amu}{\anu}cd}L_K^{\amu} L_K^{\anu} L_K^c  L_K^d, \qquad 
K_1=\ep_{{\amu}{\anu}cd}L_P^{\amu} L_K^{\anu} L_K^c  L_K^d, \qquad 
K_2=\ep_{{\amu}{\anu}cd}L_P^{\amu} L_P^{\anu} L_K^c  L_K^d, \qquad \nn\\
K_3&=&\ep_{{\amu}{\anu}cd}L_P^{\amu} L_P^{\anu} L_P^c  L_K^d, \qquad 
K_4=\ep_{{\amu}{\anu}cd}L_P^{\amu} L_P^{\anu} L_P^c  L_P^d. \qquad 
\eea 
They satisfy
\be
dK_q=(4-2q)\,K_q\,L_D\equiv (4-2q)\,\W_q,\qquad (q=0,1,2,3,4),
\ee
which counts the dilatation weight. Only $K_2$ is closed and a WZ Lagrangian is 
constructed from
\be
\W_2=K_2\,L_D=d\CL^{WZ}_2.
\ee
Other 4 invariant 5-forms $\W_q=K_q\,L_D,\; (q\neq 2)$ are closed but are CE trivial. 
There is no other possible invariant 4-form from MC forms\footnote{We may consider an invariant 4-form $(L_P^a L_{Ka})^2$ but it is exact. 
}.
Then the general 3-brane Lagrangian constructed from the MC forms is
\be
\CL=\sum_{q\neq2}\,c^q\,K_q+b^2\,\CL^{WZ}_2.
\ee
It gives the well known conformal Galileon Lagrangian 
in which the coordinate associated with $D$ is the Galileon field $\pi$.

For general odd $p$-brane there appears only one WZ action $\CL^{WZ}_{\frac{p+1}2}$
while no   WZ action exists for even $p$.  
\begin{center}
\begin{tabular}{|c c||c c|}
  \hline
 {\rm Conf}\;   & $p=3$ &  {\rm Conf}  & {\rm even}\;$p$ \\
  \hline  
   $K_{0}$  && $K_{0}$  &   \\
  $K_{1}$  && $K_1$&   \\
  & $\CL^{WZ}_{2}$  & $\vdots$&  \\
  $K_{3}$&  & $K_p $ &   \\
   $K_{4}$&  & $K_{p+1}$ &  \\
  \hline
\end{tabular}
\vs
Table 3:{~~Possible invariant and WZ Lagrangians for conformal Galileons. }
\end{center}

The variations of $K_q, (q\neq2)$ and $\CL^{WZ}_2$   with respect to  $[\D Z]_K^a$ give the common factor  $L^D$ and inverse Higgs equation
\be
L_D=0 \label{IHcondD}
\ee follows by the same reason as discussed below equation \bref{IHcond}.
Those   with respect to  $[\D Z]_P^a$ give the same common factor  $L^D$ and the EOM is satisfied identically
due to the diffeomorphism invariance.   
Finally the variation of  the action  with respect to  $[\D Z]_D$ gives 
\be
\sum_{q\neq2}\,c^q\,(2q-4)K_q+b^2\,K_2=0.
\label{confGalEOM}\ee
When we solve the IH equation \bref{IHcondD} it gives second order differential equation for
the Galileon field $\pi$, see for example in Appendix B.  

\section{Supersymmetrization} 

Galileons in the supersymmetric theories are interesting and have been examined
using four dimensional chiral superfield \cite{Khoury:2011da}
and D-brane in supergravity background \cite{Choudhury:2012yh}.
Here we apply present algebraic method of the bosonic Galileons to 
supersymmetric case. We propose a supersymmetric Galileon algebra in 5-dimensions 
and find five closed and invariant 5-forms for the WZ actions. 
They go back to the bosonic Galilean Galileon in absence of fermionic fields.  

We start from the superalgebra $su(2,2|1)$ whose bosonic subalgebra is $so(4,2)\times U(1)$ thus AdS$_5\times U(1)$. In 5D minimal spinors are symplectic Majorana U(1) doublet.
(We basically follow the spinor notations of  \cite{VanProeyen:1999ni}\cite{Freedman:2012zz}. )
The superalgebra is, using 5D Dirac matrices $\gam_A, (A=0,1,2,3,4)$, the charge conjugation matrix $C$ and Pauli matrix $\s_2$ , as
\bea
\left[M_{AB},M_{CD}\right] &=&-i \left(\h_{B[C}M_{AD]}-\h_{A[C}M_{BD]}\right),
\nn\\ 
\left[M_{AB},M_{C5}\right] &=&-i \,\h_{[BC}M_{A]5}, \qquad
\left[M_{A5},M_{C5}\right] =-i\,M_{AC}=i\h_{55}\,M_{AC},
\nn\\
 \left[M_{AB},Q^k_\B\right] &=&-\frac{i}2 \,(Q^k \gamma_{AB})_\B,
\quad 
\left[M_{A5},Q^k_\B\right] =-\frac{i}2 \,(Q \gamma_{A}\s_2)^k_\B, 
\quad  
\left[U,Q^k_\B\right] = (Q\s_2)^k_\B,\nn\\
\{ Q^i_\A,Q^k_\B\}&=&
\,M_{AB}\,(C\gamma^{AB})_{\A\B}\D^{ik}
-2\, M_{A5}(C\gamma^{A}\s_2)^{ik}_{\A\B}
-{3\,i }\,U\,(C\s_2)^{ik}_{\A\B}.
\label{SU221s}
\eea
The bosonic  generators are $M_{AB},M_{A5}, (A,B=0,1,2,3,4)$ for so(4,2) and 
$U$ for $U(1)$ and $\h_{AB}=(-;++++), \h_{55}=-1$.  
The N=2 supercharges $Q^i, (i=1,2)$ are symplectic Majorana spinors
satisfying reality condition $Q^\dagger=QB^{-1}\s_2$, $(B=C\gamma_0)$.

The bosonic generators  are rescaled for the Galilean contraction as in
\bref{Galrescale}
\be
M_{a5}=RP_{a}, \qquad M_{45}=RP_{\pi}\to {\w}R\,P_{\pi}, \qquad M_{{\amu} {4}}=B_{{\amu}}
\to \w B_{{\amu}} .
\label{Galrescales}\ee 
The supercharges $Q^i$ are rescaled to have well defined  contraction limit. 
They are divided using projection operators 
$\CP_\pm=\frac12(1\pm \gamma_4\s_2), $
as $Q^\pm=Q\CP_\pm $\footnote{Here $\gamma_4=i\gam_{0123}$ is usual $\gamma_5$ and the projections manifest so(3,1) Lorentz invariance.}. Since each $Q^+$ and $Q^-$ satisfies symplectic
Majorana  condition they can be rescaled respectively as 
 \be
Q^+\to {\sqrt{R}}\,Q^+,\qquad Q^-\to {\w\sqrt{R}}\,Q^-. 
\ee
In contrast to the bosonic case we cannot take independent limits of $R\to \infty$ and 
$\w\to \infty$
but the algebra \bref{SU221s} is contracted by 
keeping $\w/R=c\,$ a constant.  
 The contracted algebra is, rewriting $P_\amu-cM_{\amu\pi}\to P_\amu$ 
or simply choosing $c=0$,  
\bea
\left[B_{{\amu}},~P_{{\anu}}~\right]&=&i~ \h_{{\amu}{\anu}}~P_{\pi},\qquad
\left[P_{\amu},~M_{{\arho}{\as}}\right]=-i~\h_{{\amu}[{\arho}}~P_{{\as}]},
\nn\\
\left[B_{{\amu}},~M_{{\arho}{\as}}~\right]&=&-i~\h_{{\amu}[{\arho}}~B_{{\as}]},\quad
\left[M_{{\amu}{\anu}},~M_{{\arho}{\as}}\right]=-i~\h_{{\anu}[{\arho}}~M_{{\amu}{\as}]}+i~\h_{{\amu}[{\arho}}~M_{{\anu}{\as}]},
\label{AdS23s}
\eea\be
\left[B_{\amu},Q^{+ k}_\B\right]=-\frac{i}{2} \,(Q^{-k} \gamma_{\amu 4})_\B,\quad
 \left[M_{\amu\anu},Q^{\pm k}_\B\right] =-\frac{i}2 \,(Q^{\pm k} \gamma_{\amu\anu})_\B,
\quad 
\left[U,Q^{\pm k}_\B\right] =\pm(Q^\pm\gam^4)^k_\B,
\ee
\be
 \{ Q^{+i}_\A,Q^{+k}_\B\}=-{2\,}P_\amu\,(C\gamma^{\amu 4}\CP^{+})_{\A\B}^{ik}, \qquad
\{ Q^{+i}_\A,Q^{-k}_\B\}=2\, P_{\pi}(C\CP^{-} )^{ik}_{\A\B}
\label{SU221d}\ee
and other (anti-)commutators vanish. The bosonic subalgebra \bref{AdS23s} is that of the 
Galilean  Galileon (case 1). It includes supersubalgebra whose generators are  $(P_a,M_{ab},Q^{+i}_\A)$ forming a ${\cal N}=1$ superPoincar\'e algebra in four dimensions.  Note the projected symplectic Majorana supercharge  $Q^{+i}_\A,(i=1,2, $ $\A=1,...,4)$ has
$4$ real degrees of freedom. 
Although the superalgebra \bref{AdS23s}-\bref{SU221d} is a supersymmetric extension of the bosonic Galilean algebra \bref{AdS23s} it is not sufficient to obtain invariant closed forms using the non-linear realization of supercoset $G/H$ shown as below. 
It further requires two extensions, 
one is to add fermionic charge $\Sig^{-k}$ in the commutator of $\left[P,Q\right]$ as was done in case of superstring \cite{Green:1989nn},
\be
\left[P_{{\amu}},Q^{+k}_\B\right]=-\frac{i}2(\Sig^{-k} \gamma_{{\amu} 4})_\B,\quad 
 \left[M_{\amu\anu},\Sig^{- k}_\B\right] =-\frac{i}2 
\,(\Sig^{-k} \gamma_{\amu\anu})_\B,
\quad 
\left[U,\Sig^{-k}_\B\right] =-(\Sig^{-k}\gam^4)_\B.
\label{SU22dexS}\ee
The other is  two central charges $Z,\7Z$ added  in the anti-commutators. 
The second of \bref{SU221d} is replaced by
\bea
\{ Q^{+i}_\A,Q^{-k}_\B\}&=&
2\, P_{\pi}(C\CP^{-} )^{ik}_{\A\B}+
\, Z (C\C\gam^4\CP^{-} )^{ik}_{\A\B}
, 
\nn\\
\{ Q^{+i}_\A,\Sig^{-k}_\B\}&=&
\, \7Z (C\C\gam^4\CP^{-} )^{ik}_{\A\B}.
\label{SU22dex}\eea

The left invariant MC form
\be
\W= 
P_{\amu} L_P^{\amu}+\frac12\,M_{{\amu}{\anu}}\,L^{{\amu}{\anu}}+P_\pi L^\pi
+\,B_{{\amu}}\,L_B^{{\amu}} +Q^+_\A L_+^\A+Q^-_\A L_-^\A
+\Sig^-_\A L_{\Sig-}^\A+
Z\,L_Z+\7Z L_{\7Z}
\ee 
of this algebra satisfies MC equation $d\W+i\W\wedge\W=0$, 
\bea
dL_P^{{\amu}}&+&{L^{{\amu}}}_{\arho} {L}_P^{\arho} +
i ( \ba L_+^{i}\,\gamma^{{\amu} 4}\,L_+^{i}) =0,\qquad
dL^{{\amu}{\anu}}+L^{{\amu}{\arho}}{L_{\arho}}^{\anu}=0,
\nn\\
dL^{\pi}&-&{L_{B\arho}} {L}_P^{\arho}-
{2i} ( \ba L_+^{i}\,L_-^{i}) =0,\qquad
dL_B^{{\amu}}+{L^{{\amu}}}_{{\arho}}{L_B^{{\arho}}}=0,
\qquad
dL_U=0,
\nn\\
dL_+^{i\A}&+&\frac14\,L^{{\amu}{\anu}}(\gamma_{{\amu}{\anu}}{L_+})^{i\A}
+i\,L_U\,(\gam^4\,L_+)^{i\A}=0,
\nn\\
dL_-^{i\A}&+&\frac14\,L^{{\amu}{\anu}}(\gamma_{{\amu}{\anu}}{L_-})^{i\A}
-i\,L_U\,(\gam^4\,L_-)^{i\A}
+\frac1{2}\,L_B^{{\amu}}(\gamma_{{\amu} 4}{L_+})^{i\A}
=0,
\nn\\
dL_{\Sig-}^{i\A}&+&\frac14\,L^{{\amu}{\anu}}(\gamma_{{\amu}{\anu}}{L_{\Sig-}})^{i\A}
-i\,L_U\,(\gam^4\,L_{\Sig-})^{i\A}
+\frac1{2}\,L_P^{{\amu}}(\gamma_{{\amu} 4}{L_+})^{i\A}
=0,\nn\\
dL_Z&-&{i} ( \ba L_+^{i}\,\gam^4\,L_-^{i}) =0,
\qquad
dL_{\7Z}-{i} (\ba L_+^{i}\,\gam^4\,L_{\Sig-}^{i}) =0.
\label{MCSU221e2ex}
\eea
 The set of MC equations are consistent under the operation of  "$d$" guaranteeing 
the closure of the superalgebra \bref{AdS23s}-\bref{SU22dex}.
In appendix C we present forms of $L$'s in a choice of coset parametrization though 
they are not used in the following. 

Using the superalgebra $G$ and a coset $G/H=G/(SO(3,1)\times U(1))$
we will construct invariant 4-and 5-forms.  
$H$-invariant and closed 4-forms $\7K_q, (q=0,...4)$ which are reduced to the bosonic
$K_q$ in \bref{defKq} are  
\bea
\7K_0&=&\ep_{{\amu}{\anu}{\arho}{\as}}L_B^{{\amu}} L_B^{{\anu}}L_B^{{\arho}}
L_B^{{\as}},
\nn\\
\7K_1&=&\ep_{{\amu}{\anu}{\arho}{\as}}\{ L_P^{{\amu}} L_B^{{\anu}}-i\,(\ba L_+\gam^{{\amu}{\anu}}L_-)\}L_B^{{\arho}}L_B^{{\as}}, 
\nn\\ 
\7K_2&=&\ep_{{\amu}{\anu}{\arho}{\as}}\{ L_P^{{\amu}} L_B^{{\anu}}-i\,(\ba L_+\gam^{{\amu}{\anu}}L_-)\}
\{ L_P^{{\arho}} L_B^{{\as}}-i\,(\ba L_+\gam^{{\arho}{\as}}L_-)\},
\nn\\
\7K_3&=&\ep_{{\amu}{\anu}{\arho}{\as}}\left\{ L_P^{{\amu}} L_P^{{\anu}}
-2i\, (\ba L_+\gam^{{\amu}{\anu}}L_{\Sig-})\right\}\{ L_P^{{\arho}} L_B^{{\as}}-i\,(\ba L_+\gam^{{\arho}{\as}}L_-)\},
\nn\\ 
\7K_4&=&\ep_{{\amu}{\anu}{\arho}{\as}}
\left\{ L_P^{{\amu}} L_P^{{\anu}}
-2i\, (\ba L_+\gam^{{\amu}{\anu}}L_{\Sig-})\right\}\left\{ L_P^{{\arho}} L_P^{{\as}}
-2i\, (\ba L_+\gam^{{\arho}{\as}}L_{\Sig-})\right\}
\label{tildeKq}\eea
and 
\be
d\,\7K_q=0,\qquad {(q=0,1,...,4)}.
\ee
Here in order to construct closed  $\7K_3$ and  $\7K_4$ we need to introduce $L_{\Sig-}$ 
associated to the supercharge ${\Sig^-}$ added in \bref{SU22dexS}. 

Similarly closed and invariant 5-forms  which are reduced to the bosonic
$\W_q$ in \bref{defWq}  are constructed as 
\bea
\7\W_0
&=&\7K_0\,L^\pi-2i\,\ep_{{\amu}{\anu}{\arho}{\as}}L_B^{{\amu}}L_B^{{\anu}}
L_B^{{\arho}}(\ba L_-\gam^{{\as} 4}L_-),
\nn\\ 
\7\W_1
&=&\7K_1\,L^\pi-2i\,\ep_{{\amu}{\anu}{\arho}{\as}}\,L_B^{{\amu}}
\left\{L_B^{{\anu}}L_B^{{\arho}}-i (\ba L_+\gam^{{\anu}{\arho}}L_-)\right\}
(\ba L_-\gam^{{\as} 4}L_-), 
\nn\\
\7\W_2&=&\7K_2\,L^\pi-2i\,\ep_{{\amu}{\anu}{\arho}{\as}}
\,L_P^{{\amu}}\,\left( L_P^{{\anu}}L_B^{{\arho}}-2i
(\ba L_+\gam^{{\amu}{\anu}}L_-)\right)(\ba L_-\gam^{{\as} 4}L_-)
\nn\\&&-16i\,L_Z\,(\ba L_+\gam^{4}L_-)\,(\ba L_+\gam^{4}L_-),
\nn\\ 
\7\W_3&=&\7K_3\,L^\pi
-i\,\ep_{{\amu}{\anu}{\arho}{\as}}
\,L_P^{{\amu}}\,\left\{ L_P^{{\anu}}L_P^{{\arho}}-4i (\ba L_+\gam^{{\anu}{\arho}}L_{\Sig-})\right\}
(\ba L_-\gam^{{\as} 4}L_-)
\nn\\
&-& 2i\,\ep_{{\amu}{\anu}{\arho}{\as}}
\, L_P^{{\amu}}\left\{ L_P^{{\anu}}L_B^{{\arho}}-3i (\ba L_+\gam^{{\anu}{\arho}}L_{-})\right\}(\ba L_-\gam^{{\as} 4}L_{\Sig-})+
\nn\\
&+&i\,\ep_{{\amu}{\anu}{\arho}{\as}}
L_B^{{\amu}}\left\{  L_P^{{\anu}}L_B^{{\arho}}-i (\ba L_+\gam^{{\anu}{\arho}}L_{-})\right\}(\ba L_{\Sig-}\gam^{{\as} 4}L_{\Sig-})
\nn\\
&+&\frac13\,\ep_{{\amu}{\anu}{\arho}{\as}}
\left\{2 L_P^{{\amu}}(\ba L_-\gam^{{bcd}}L_{\Sig-})+ 
 L_B^{{\amu}}(\ba L_{\Sig-}\gam^{{bcd}}L_{\Sig-})\right\}
(\ba L_{+}\gam^{4}L_{-})
\nn\\&-&32i\,L_{\7Z}\,(\ba L_+\gam^{4}L_-)\,(\ba L_+\gam^{4}L_-),
\nn\\ 
\7\W_4&=&\7K_4\,L^\pi
-4i\,\ep_{{\amu}{\anu}{\arho}{\as}}\,L_P^{{\amu}}
\left\{L_P^{{\anu}}L_P^{{\arho}}-4i(\ba L_{+}\gam^{{\anu}{\arho}}L_{\Sig-})\right\}
(\ba L_{-}\gam^{{\as} 4}L_{\Sig-})
\nn\\&+&2i\,\ep_{{\amu}{\anu}{\arho}{\as}}\,L_P^{{\amu}}
\left\{L_P^{{\anu}}L_B^{{\arho}}+2i(\ba L_{+}\gam^{{\anu}{\arho}}L_{-})\right\}
(\ba L_{\Sig-}\gam^{{\as} 4}L_{\Sig-})
\nn\\&+&8\,\ep_{{\amu}{\anu}{\arho}{\as}}\,L_B^{{\amu}}
(\ba L_{+}\gam^{{\anu}{\arho}}L_{\Sig-})(\ba L_{\Sig-}\gam^{{\as} 4}L_{\Sig-})
+\frac43\,\ep_{{\amu}{\anu}{\arho}{\as}}
L_P^{{\amu}}(\ba L_{+}\gam^{4}L_{-})(\ba L_{\Sig-}\gam^{{\anu\arho\as} }L_{\Sig-})
\nn\\&-&64i\,L_{\7Z}\,(\ba L_+\gam^{4}L_-)\,(\ba L_+\gam^{4}L_{\Sig-}),
\label{Wqsuper}\eea
and 
\be
d\,\7\W_q=0,\qquad {(q=0,1,...,4)}.
\ee   
In order to have closed 5-form $\7\W_2$ we need the central charge $Z$ 
and  to get $\7\W_3$ and $\7\W_4$ we use the central charge $\7Z$
 as well as $\Sigma^-$.  

These $\7K_q$ and $\7\W_q$  go back to the bosonic ones $K_q$ in \bref{defKq} and $\W_q$ in \bref{defWq} when the fermions are put to zero and 
are the supersymmetric extensions of  the Galilean Galileon  in \bref{GalileondK}. 
All $\7K_q, (q=0,1,2,3,4)$ are closed and are surface term. 
All invariant closed 5-forms $\7\W_q, (q=0,1,2,3,4)$ are CE non-trivial since the bosonic pieces are non-trivial.    
Thus  $\7\W_q, (q=0,1,2,3,4)$ are used to construct five WZ Lagrangians of the supersymmetric model. 
 
If we restrict ones which have bosonic body they are unique invariants, up to surface terms,  as in the bosonic case. 
However there are other $H$-invariant, thus $G$-invariant,  4 and 5-forms which vanish when fermions are put to zero.    
For example a piece in $\7K_1$ in \bref{tildeKq}
\be
\ep_{{\amu}{\anu}{\arho}{\as}}\,i\,(\ba L_+\gam^{{\amu}{\anu}}L_-)L_B^{{\arho}}L_B^{{\as}},  
\ee
is  $H$-invariant 4-form. 
There are number of such invariant fermionic  Lagrangians  that could be 
 added to the Lagrangian consistent with the supersymmetry.

\section{Summary and Discussions} 

In this paper we have reexamined the cohomological structure of the Galileon models
using MC equations of the Galileon algebras and understood how the Lagrangians appear as
invariant 4-forms and/or pseudo invariant WZ  terms. 
As we can write the EOM in terms of MC forms we can understand why the 
inverse Higgs condition $L^\pi=0$ appears from the $H$-covariance.
It also manifests that the Galileon scalar $\pi$ satisfies second order EOM.  
It is noticed that they are shown to hold without using particular parametrizations of the coset.  
\vs

We have also proposed a supersymmetric Galileon algebra that contains 
bosonic Galileon algebra and the ${\cal N}=1$ superPoincar\'e algebra as its subalgebras.  
We have constructed supersymmetric counterparts of the invariant and closed 4-forms
and  5-forms of the Galilean Galileon. The former are surface term and the latter are used to construct the WZ Lagrangians. 
If we restrict ones which have bosonic body, that does not vanish 
when fermions are put to zero, they are unique ones.
However there is an ambiguity of  $H$-invariant  4-forms which vanish when fermions are put to zero.    

There are several issues to be discussed further for establishing the supersymmetric Galileon
theory.  
In constructing the supersymmetric Lagrangians we take pullback of the MC forms. 
There are two options one is pullback to 4-dimensional Minkowski space with coordinates
$x^\mu$ and other is  pullback to ${\cal N}=1$ superspace with coordinates
$(x^\mu,\T^\A_+)$.  In the former case fields appears as 
\be
\pi(x), \; v^a(x), \; \T_\pm(x),....
\ee
In the latter case 
fields appear as superfields,
\be
\pi(x,\T_+),\; v^a(x,\T_+), \; \T_-(x,\T_+), \;  .... 
\ee
The superfield $\pi(x,\T_+)$ when expanded by fermionic coordinates $\T_+$ defines the Galileon 
supermultiplet. The  (super)transformations are non-linearly realized on these fields 
following to \bref{inftransg}.  
It is necessary to write down the Lagrangian and clarify nature of 
dynamical fields and auxiliary fields.  

In the bosonic case the inverse Higgs condition $L^\pi=0$ is derived 
 as the EOM. 
Although it is concluded from the covariance in the bosonic case,
it is not clear for the supersymmetric case since we can construct fermion bi-linear covariants
which could appear in the variations. It is important that the EOM  is solved for   the broken boost variables $v^{a}$ algebraically for the Galileon scalar
satisfying second order EOM. 
(There is an option to impose 
  the   inverse Higgs condition\cite{Ivanov:1975zq}  to reduce the  boost variables $v^{a}$.)

If the model is considered as a relativistic 3-brane it is natural that the supersymmetric 
model possess kappa symmetries.  
However the 3-brane in 4-dimensions is not dynamical, filled in whole space-time,  and the superGalileon appears as supersymmetric field theory in 4-dimensions the role of kappa symmetries is not evident.   
 Both the kappa invariance and the appearance of IH condition depend 
on the choice of Lagrangian. It is interesting to examine 
if 
the kappa symmetry can be satisfied by fixing the above mentioned  ambiguity 
of fermionic $H$-invariant Lagrangian terms.

It is also interesting to make clear the relation to other approaches 
of the supersymmetric Galileons. For example in \cite{Khoury:2011da} four 
dimensional (conformal) Galileon Lagrangians are supersymmetrized using 
chiral superfield, while we consider ones from reduction of five dimensional algebra.
As bosonic Galileons are understood from higher dimensions 
we expect the superGalileon is derived in the same way naturally.  
These remaining issues are discussed in future investigations. 

\vs
 {\bf Acknowledgements }  
 
The authors would like to thank Joaquim Gomis, Yoshikane Honda and Erika Takeda for 
valuable discussions.
\appendix
\section{General properties}

Here we consider general  graded Lie algebra,
\be
[G_A,G_B\}=if^C_{AB}\,G_C, 
\qquad f^C_{AB}=-(-)^{AB}f^C_{BA},
\ee
where we use $A,B...$ for even and odd generators of $G$. 
$(-)^{AB}=-1$ only when both $G_A$ and $G_B$ are odd. The MC one form is
\be
\W=-ig^{-1}dg=G_AL^A,
\ee
and MC equation is 
\be
d\W+i\W^2=0,\qquad dL^A+\frac12f^A_{BC}L^CL^B=0.
\label{MCeqL}\ee
The consistency is equivalent to hold the Jacobi identity
\bea
0&=&f^A_{BC}\{(dL^C)L^B-L^C(dL^B)\} 
\too 
f^A_{BC}\,f^C_{DE}\,L^E\,L^D\,L^B=0.
\eea
Using  coset coordinates  $Z^M$  of the $G/H$, the MC form components $L^A$ are expressed as $L^A=dZ^M{L_M}^A(Z)$,  and the MC equation becomes
\be
dL^A+\frac12f^A_{BC}L^CL^B=-dZ^MdZ^N\pa_N{L_M}^A+
\frac12f^A_{BC}dZ^M{L_M}^CdZ^N{L_N}^B=0,
\ee
where $\pa_M$ is the left derivative  with respect to  $Z^M$.   Then it holds
\be
\pa_M{L_N}^A-(-)^{MN}\pa_N{L_M}^A+
\frac12f^A_{BC}\left((-)^{NC}{L_M}^C{L_N}^B-(-)^{MN+MC}{L_N}^C{L_M}^B\right
)=0.
\label{A6}\ee
We define $[\D Z]^A $ by replacing $dZ^M$ with $\D Z^M$ in $L^A$,
\be
[\D Z]^A =\D Z^M{L_M}^A.
\ee
Using \bref{A6} the variation of $L^A$ , under {\it any}  variation $\D Z^M$, is computed as 
\bea
\D L^A&=&(d\D Z^M){L_M}^A+dZ^M \D Z^N\pa_N{L_M}^A
= d[\D Z]^A +f^A_{BC}\;{L}^C{[\D Z}]^B.
\label{deltaLIF}\eea
Remember this formula holds for any graded algebras in this ordering.  

 \vs
 \section{Explicit parametrization of coset } 
 
 Here we solve the IH equation $L^\pi=0$ in \bref{IHcondsol} and express MC forms in terms of Galileon fields. 
 
The explicit form of the MC form $L$'s depends 
on the parametrization of the coset. We parametrize the coset $G/SO(p,1)$, for example, as
\be
g=e^{iP_{\amu} x^{\amu}}\,e^{iP_\pi \pi}\, e^{iM_{{\amu}\pi}v^{\amu}}=
g_0\,e^{iP_\pi \pi}\, e^{iM_{{\amu}\pi}v^{\amu}},\qquad g_0\equiv e^{iP_{\amu} x^{\amu}}.
\ee
We first compute $\W_0=-ig_0dg_0,$   
\be
-ig_0^{-1}dg_0=-i\, e^{-iP_{\amu} x^{\amu}}d e^{iP_{\amu} x^{\amu}}=
e^{\amu}\,P_{\amu}+\frac12\w^{{\amu}{\anu}}\,M_{{\amu}{\anu}},
\ee with
\be
e^{\amu}=dx^{\amu}+{O(x)^{\amu}}_{\anu} dx^{\anu} \,(\frac{\sh(\frac{X}{R})}{\frac{X}{R}}-1),
\quad
\w^{{\amu}{\anu}}=\frac{dx^{[{\amu}}x^{{\anu}]}}{R^2}\,\frac{(\ch(\frac{X}{R})-1) }{(\frac{X}{R})^2},
\ee
where $X=\sqrt{x_{\amu} x^{\amu}}$ and ${O_{\amu}}^{\anu}(x) ={\D_{\amu}}^{\anu}-\frac{x_{\amu} x^{\anu}}{X^2}$. 
 $e^{\amu}$ and $\w^{{\amu}{\anu}}$ are viel-bein and spin connection verifying
 the AdS MC equations,
\be
de^{\amu}+\w^{{\amu}{\anu}}e_{\anu}=0,\qquad d\w^{{\amu}{\anu}}+\w^{{\amu}{\arho}}{\w_{\arho}}^{{\anu}}+
\frac1{R^2}
e^{\amu} e^{\anu}=0.
\ee  
The full left invariant MC 1-forms are 
\bea
L^{\amu}&=&
\left(e^{\amu}+{\frac{v^{\amu}(e^{\anu} v_{\anu})}{V^2}}(\cos(\frac{  V}{\w})-1)
\right)\ch(\frac{\pi}{R\w})
-d\pi\,\frac{v^{\amu}}{\w V}\,\sin(\frac{ V}{\w}), 
\nn\\
L^{{\amu}{\anu}}&=&
\w^{{\amu}{\anu}}\,+
\frac{Dv^{[{\amu}}v^{{\anu}]}}{V^2}(\cos(\frac{  V}{\w})-1)-\,
e^{[{\amu}}v^{{\anu}]}\frac{ 1}{R V}\,\sin(\frac{  V}{\w})\,\sh(\frac{\pi}{R\w}),
\nn\\
L^\pi&=&
d{\pi}\,\cos(\frac{ V}{\w}) +
(e^{\anu} v_{\anu})\,\frac{\w}{V}\,\sin(\frac{  V}{\w})\ch(\frac{\pi}{R\w}) ,
\label{MCgeneral} \\
L^{{\amu}\pi}&=&
 Dv^{\amu}+
{O^{\amu}}_{\anu}(v)\,Dv^{\anu}\,(\frac{ \w}{ V}\sin(\frac{V}{\w})-1) +
\left(e^{\amu}+{O^{\amu}}_{\anu}(v)e^{\anu} \,(\cos(\frac{  V}{\w})-1)\right)\,\frac{\w}{R}\,
\sh(\frac{\pi}{R\w}),
\nn\eea
where
$V=\sqrt{v_{\amu} v^{\amu}},\,{O_{\amu}}^{\anu}(v) ={\D_{\amu}}^{\anu}-\frac{v_{\amu} v^{\anu}}{V^2}$
and $Dv^{\amu}=dv^{\amu}+\w^{{\amu}{\anu}}v_{\anu}$. 
\vs

We solve the Higgs constraint for each Galileon cases.
 \vs

\noindent{\underline{ 1) Galilean brane (Galileon)}}  [$\w\to\infty, R\to\infty$] 
\vs
The Higgs equation \bref{IHcondsol}  is solved as,   
\be
L^\pi=d{\pi}\,+dx^a v_a =d\s^\mu(\pa_\mu\pi+{e_\mu}^a\,v_a)=0,\too 
v_a=-{e_a}^\mu\pa_\mu\pi,
 \ee
 where $\s^\mu, (\mu=0,1,2,3)$ are parameters of the 3-brane,
 ${e_\mu}^a={\pa_\mu x}^a$ is viel-bein and  
 ${e_a}^\mu$ is its inverse.   Using it
\bea
L^a&=& dx^a=d\s^\mu\,{e_\mu}^a, \qquad L^{ab}=\w^{ab}=0.
\nn\\
L^{a\pi}&=&dv^a= d\s^\mu \pa_\mu v^a=
-d\s^\mu\, {e^a}^\nu\nabla_\nu\nabla_\mu\pi,
\eea
where the covariant derivative $\nabla_\mu$ is with respect to the induced metric
$g_{\mu\nu}=\pa_\mu x^a\pa_\nu x^b\h_{ab}$. %
 If we take a  static gauge $x^a=\s^a, \;{e_\mu}^a={\D_\mu}^a$ it becomes
 \be
 v_a=-\pa_a\pi,\qquad L^a= dx^a, 
\qquad
L^{a\pi}=
-dx^b\, \pa^a\pa_b\pi.
 \ee
{\underline{ 2) Poincar\'e brane (DBI Galileon)}}   [$\w=1, R\to\infty$] , 
\vs
In this limit the Higgs equation   is solved as, 
\bea
L^\pi&=&
d{\pi}\,\cos(V) +(e^a v_a)\,\frac{\sin V}{V}=0,\,\too 
\7v_a\equiv v_a\,\frac{\tan V}{V}=-{e_a}^\mu\pa_\mu\pi.
 \eea
 where $V=\sqrt{v^2},\;\7V=\sqrt{\7v^2}=\tan V,\; (\pa\pi)^2=g^{\mu\nu}\pa_\mu\pi\pa_\nu\pi$ and 
\bea
L^a&=&\left(e^{\amu}+{\frac{v^{\amu}(e^{\anu} v_{\anu})}{V^2}}(\cos {  V}-1)
\right) 
-d\pi\,\frac{v^{\amu}}{V}\,\sin V
\nn\\&=&
d\s^\nu e^{a\mu}\left(g_{\mu\nu}+{\frac{\pa_\mu\pi \pa_\nu\pi}{(\pa\pi)^2}}({\sqrt{1+(\pa\pi)^2}}-1)\right), 
\eea
\bea
L^{a\pi}&=&dv^a+
{O^a}_b(v)\,dv^b\,(\frac{ \sin V}{ V}-1)
\nn\\&=&
d\s^\mu{e^{a\nu}}
\left(-\frac{\nabla_\nu\nabla_\mu\pi}{\sqrt{1+{(\pa\pi)^2}}}+
\frac{\pa_\nu\pi \pa_\mu(\pa\pi)^2}{2(\pa\pi)^2({1+{(\pa\pi)^2}})}
(\sqrt{{1+{(\pa\pi)^2}}}-1)\right).
\eea 
In the static gauge $x^a=\s^a, \;{e_\mu}^a={\D_\mu}^a$ and $\nabla_\mu=\pa_\mu$. 
\vs
{\underline{ 3)  NH brane} }  [$\w\to\infty, R$ finite]. 
\bea
L^\pi&=&d{\pi}\,+(e^a v_a) =d\s^\mu(\pa_\mu\pi+{e_\mu}^a\,v_a)=0,\too 
v_a=-{e_a}^\mu\pa_\mu\pi,
 \nn\\
L^a&=& e^a=d\s^\mu\,{e_\mu}^a, 
 \nn\\
L^{a\pi}&=& Dv^a+e^a\,\frac{\pi}{R^2}
=-d\s^\mu{e}^{a\nu}\left(\nabla_\nu\nabla_\mu\pi-{g_{\nu\mu}}\,\frac{\pi}{R^2}\right)
\eea
where 
$ D_\mu v^a=-{e}^{a\nu}\nabla_\nu\nabla_\mu\pi$ 
and the covariant derivative $\nabla_\mu$ is  with respect to  the AdS$_4$ metric $g_{\mu\nu}={e_\mu}^a{e_\nu}^b\h_{ab}.$
 \vs 
{\underline{4) AdS  brane}   [$\w=1, R$ finite],  
\vs
For AdS Galileon $L^\pi=0$ is solved as 
\be
L^\pi=
d{\pi}\,\cos({ V}) +
\,\frac{(e^\nu v_\nu)}{V}\,\sin({  V})\ch(\frac{\pi}{R})=0,\,\too 
\7v_a\equiv v_a\,\frac{\tan { V}}{V}=-{e_a}^\mu\pa_\mu\Pi, 
 \ee
where
$\;\tan(\frac{\Pi}{2R})=\tanh(\frac{\pi}{2R}),\; $ and
\bea
L^a&=&
d\s^\nu e^{a\mu}
\left({g_{\mu\nu}}+{\frac{\pa_\mu\Pi \pa_\nu\Pi}{(\pa\Pi)^2}}({\sqrt{1+(\pa\Pi)^2}}-1) 
\right)\frac1{\cos(\frac{\Pi}{R})}, \qquad (\pa\Pi)^2=g^{\mu\nu}{\pa_\mu\Pi \pa_\nu\Pi}
\nn\\
L^{a\pi}&=&
d\s^\mu{e^{a\nu}}
\left(-\frac{\nabla_\nu\nabla_\mu\Pi}{\sqrt{1+{(\pa\Pi)^2}}}+
\frac{\pa_\nu\Pi \pa_\mu(\pa\Pi)^2}{2(\pa\Pi)^2({1+{(\pa\Pi)^2}})}
(\sqrt{1+{(\pa\Pi)^2}}-1)\right)\nn\\&+&
d\s^\mu{e^{a\nu}}
\left(\frac{{g_{\mu\nu}}}{\sqrt{1+{(\pa\Pi)^2}}}-\frac{\pa_\mu\Pi\pa_\nu\Pi}{(\pa\Pi)^2}
(\frac{1}{\sqrt{1+{(\pa\Pi)^2}}}-1)\right)\,\frac{\tan(\frac{\Pi}{R})}{R}.
\eea
In $R\to\infty$ and/or  $\w\to\infty$ limits they go to ones of contracted results. 

\vs 
For {\underline{ the conformal Galileon}}  in section 4 
 we parametrize the coset $G/SO(p,1)$ as
\be
g=e^{iP_a x^a}\,e^{iD \pi}\,e^{iK_a v^a}\,
\ee
the MC forms are 
\bea
 L_P^{a}&=&  e^\pi\,d x^a,\qquad 
 L_K^{\amu}=dv^{{\amu}}+v^a\;d\pi-v^2\, e^\pi dx^{\amu}+2\, e^\pi\,v^{\amu}\,(vdx),\nn\\
 L_D&=&d\pi+2\,e^\pi\,(vdx),\qquad L^{{\amu}{\anu}}=2\,e^\pi\,v^{[{\amu}}dx^{{\anu}]}. 
\eea
Solving $L_D=0$ as
\be
v^a=-\frac12{e_a}^{\amu}\pa_{\amu}\pi,\qquad {e_\mu}^a=e^\pi\,\pa_\mu x^a,
\ee $v^a$ is eliminated in terms of the conformal Galileon field $\pi$ and 
\be
 L_P^{a}= d\s^\mu\,  {e_\mu}^a,\qquad 
 L_K^{\amu}=-\frac12 d\s^\mu\,  {e^{a\nu}}\left(\nabla_\mu\nabla_\nu\pi+\frac12g_{\mu\nu}
(\pa\pi)^2\right), \label{Lconf2}\ee
where $\nabla_\mu$ is  with respect to  the conformal metric $g_{\mu\nu}={e_\mu}^a{e_\nu}^b\h_{ab}$.
The EOM for the Galileon $\pi$ is given from \bref{confGalEOM} using \bref{Lconf2}. 
\vs
 
 \section{Supercoset $G/(SO(3,1)\times U(1))$} 
 
The explicit form of the MC form $L$'s depends on the parametrization of the coset. 
For the supercoset  $G/(SO(3,1)\times U(1))$ for the superalgebra $G$ in \bref{AdS23s}-\bref{SU22dex} we parametrize the coset element  $g$, for example,  as 
\be
g=e^{iP_ax^a}\,e^{iP_\pi \pi}\, e^{iQ^+\T_+}\,e^{iQ^-\T_-}\,e^{i\Sig^-\xi_-}\, e^{iB_{a}v^a}\,e^{iZc}\,e^{i\7Z\7c}.
\ee
where $\T_\pm$ and $\xi_-$ are odd coordinates of the coset associated to 
$Q^\pm$ and $\Sig^-$ and $c,\7c$ are even scalar coordinates of the central charges $Z,\7Z$.
The left invariant MC form $\W=-g^{-1}dg$ is computed as 
\be
\W= 
P_{\amu} L_P^{\amu}+\frac12\,M_{{\amu}{\anu}}\,L^{{\amu}{\anu}}+P_\pi L^\pi
+\,B_{{\amu}}\,L_B^{{\amu}} +Q^+_\A L_+^\A+Q^-_\A L_-^\A
+\Sig^-_\A L_{\Sig-}^\A+Z\,L_Z+\7Z L_{\7Z},
\ee 
where 
\bea
L_P^a&=& dx^a-i \,\ba\T_+^i \gam^{a4}d\T_+^i, 
\qquad L_B^{a}=dv^a, \qquad L^{ab}=0,\qquad L_U=0,
\nn\\
L^\pi&=& d\pi  -2i\,\ba\T_-^id\T_+^i+ v_a (dx^a
-i \, \ba\T_+^i \gam^{a4}d\T_+^i),
\nn\\
L_+^i&=&d\T_+^i, \qquad 
L_-^i=d\T_-^i-\frac12 \gam_{a4}v^a d\T_+^i,
\qquad  
L_{\Sig-}^i=d\xi_-^i+\frac12 \gam_{a4}\T_+^i\, dx^a,
\nn\\ 
L_Z&=&dc+i\ba\T_-^i\gam^{4}d\T_+^i,\qquad L_{\7Z}=d{\7c}+i\ba\xi_-^i\gam^{4}d\T_+^i.
\label{WWs4}\eea
They satisfy the MC equations \bref{MCSU221e2ex} and are building blocks of the 
invariant forms $\W_q$ \bref{Wqsuper} for superGalileon WZ Lagrangians.


\end{document}